\documentclass[11pt]{article}
\usepackage{moriond,epsfig}

\def\gev{\, \mbox{GeV}}
\def\tev{\, \mbox{TeV}}
\def\KL{K\"all\'en-Lehmann}
\def\Journal#1#2#3#4{{#1} {\bf #2}, #3 (#4)}

\def\NPB{{\em Nucl. Phys.} B}
\def\PLB{{\em Phys. Lett.}  B}
\def\PRL{\em Phys. Rev. Lett.}
\def\PRD{{\em Phys. Rev.} D}
\def\ZPC{{\em Z. Phys.} C}
\def\IJMA{{\em Int. J. Mod. Phys} A}
\def\be{\begin{equation}}
\def\ee{\end{equation}}
\def\bea{\begin{eqnarray}}
\def\eea{\end{eqnarray}}

\begin{document}
\vspace*{4cm}
\title{SM SCALAR AND EXTRA SINGLET(S)}

\author{ J. J. VAN DER BIJ }

\address{Institut f\"ur Physik, Albert-Ludwigs Universit\"at Freiburg, H. Herderstr. 3,\\
79104 Freiburg i.B., Deutschland}

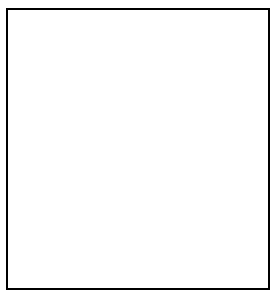
\maketitle\abstracts{I discuss the question whether it is possible that the LHC
will find no signal for the Higgs particle. It is argued that in this case singlet scalars
should be present that could play an important role in astroparticle physics.
A critical view at the existing electroweak data shows that this possibility
might be favored over the simplest standard model. In this case one needs the ILC
in order to study the Higgs sector.
}

\section{Introduction}
The standard model gives a good description of the bulk of the electroweak data.
Only a sign of the Higgs particle is  missing at the moment. The Higgs field is
necessary in order to make the theory renormalizable, so that predictions are
possible and one can really speak of a theory. A complete absence of the 
Higgs field would make the theory non-renormalizable, implying the existence
of new strong interactions at the TeV scale. Therefore one is naively led to the
so-called no-lose theorem~\cite{chanowitz}. This theorem says that when one
 builds a large energy
hadron collider, formerly the SSC now the LHC, one will find new phyics, either 
the Higgs particle or otherwise new strong interactions. Since historically
no-theorems have a bad record in physics one is naturally tempted to try
to evade this theorem. So in the following I will try to find ways by which
the LHC can avoid seeing any sign of new physics. 

At the time of the introduction of the no-lose theorem very little was known
about the Higgs particle. Since then there have been experiments at LEP, SLAC and
the Tevatron, that give information on the Higgs mass. Through precise measurements
of the W-boson mass and various asymmetries one can get constraints on the
Higgs mass. The Higgs mass enters into the prediction of these quantities via radiative
corrections containing a virtual Higgs exchange. Moreover at LEP-200 the direct search
gives  a  lower limit of $114.4 \gev$. The situation regarding the precision tests 
is not fully satisfactory. The reason is that the Higgs mass implied by the 
forward-backward asymmetry $A_{FB}(b)$ from the bottom quarks is far away from the
mass implied by the other measurements, that agree very well with each other.
No model of new physics appears to be able to explain the difference.
From $A_{FB}(b)$ one finds $m_H= 488^{+426}_{-219} \gev$ with a $95\%$ lower bound
of $m_H=181 \gev$. Combining the other experiments one finds
$m_H= 51^{+37}_{-22} \gev$ with a $95\%$ upper bound of $m_H=109 \gev$. The $\chi^2$
of the latter fit is essentially zero. Combining all measurements gives a bad fit.
One therefore has a dilemma. Keeping all data one has a bad fit. Ignoring the 
$b$-data the standard model is ruled out. In the last case one is largely forced 
towards the extended models that appear in the following. Accepting a bad fit  one has
somewhat more leeway, but the extended models are still a distinct possibility.

\section{Is a very heavy Higgs boson possible?}
One way to avoid seeing the Higgs boson would be if it is too heavy to
be produced at the LHC.
At first sight this possibility appears to be absurd given the precision data.
Even if one takes all data into account there is an upper limit
of $m_H=190 \gev$. However the question is surprisingly difficult to answer
in detail. The reason is that the Higgs mass is not a free parameter in the 
Lagrangian. Because of the spontaneous symmetry breaking the Higgs mass
is determined by  its self-coupling $\lambda$ and the vacuum expectation value
$f$: $m^2_H=\lambda f^2$. This means that  a heavy Higgs boson is strongly interacting.
Therefore higher-loop effects can become important. These effects give corrections to
the precision measurements with a behaviour $m_H^{2.(loop-1)}$. These effects
can in principle cancel the one-loop $log(m_H)$ corrections, on which the 
limits are based. Therefore one could have the following situation: the 
strong interactions compensate for the loop effects, so that from the
precision measurements the Higgs appears to have a mass of $50 \gev$.
At the same time the Higgs is so heavy that one does not see it at the LHC.
For this to happen the Higgs mass would have to be about $3 \tev$.
Detailed two-loop~\cite{bij1,bij2,bij3,jikia,ghinculov} and 
non-perturbative $1/N$ calculations~\cite{binoth1,binoth2} have shown
that the first important effects are expected at the three-loop level.
The important quantity is the sign of the three-loop correction compared
to the one-loop correction. This question was settled in a large calculation
that involved of the order of half a million Feynman diagrams~\cite{boughezal1,boughezal2}.
The conclusion is that the strong interactions enhance the effects of a heavy Higgs
boson. This conclusion is confirmed by  somewhat
qualitative non-perturbative estimates~\cite{kastening,akhoury}. 
Therefore the Higgs boson cannot be too heavy to be seen at the LHC.

\section{Singlet scalars}
\subsection{Introduction}
If the Higgs boson is not too heavy to be seen the next try to make it invisible
at  the LHC is to let it decay into particles that cannot be detected.
For this a slight extension of the standard model is needed. In order not to effect
the otherwise good description of the electroweak data by the standard model one introduces singlet
scalars. The presence of singlets will not affect present electroweak phenomenology
in a significant way, since their effects in precision tests appear first at the two-loop
level and are too small to be seen~\cite{kyriazidou}. These singlet scalars will not couple to ordinary
matter in a direct way, but only to the Higgs sector. It is acually quite natural to expect
singlet scalars to be present in nature. After all we know there also exist singlet fermions,
namely the right handed neutrino's. The introduction of singlet scalars affects the 
phenomenology of the Higgs boson in two ways. On the one hand one creates the possibility
for the Higgs boson to decay into said singlets, on the other hand there is the possibility
of singlet-doublet mixing, which will lead to the presence of more Higgs bosons however with
reduced couplings to ordinary matter. In the precision tests this only leads to the replacement
of the single Higgs mass by a weighted Higgs mass and one cannot tell the difference between the 
two cases. Mixing and invisible decay can appear simultaneously. For didactical purpose I show
in the following simple models consisting of pure invisible decay or pure mixing. For a mini-review
of the general class of models see ref.~\cite{bij2006}.

\subsection{Invisible decay}

When singlet scalars are present it is possible that the Higgs boson decays
into these  scalars if they are light enough. 
Such an invisible  decay is rather natural, when one introduces the Higgs singlets $S_i$ as 
multiplets of a symmetry group~\cite{ste1,ste2,ste3,ste4,ste5,ste6}, for instance $O(N)$.
 When the $O(N)$ symmetry group
stays unbroken this leads to an invisibly decaying Higgs boson through
the interaction $\Phi^{\dagger}\Phi S_i S_i$, after spontaneous breaking of the
standard model gauge symmetry. 
When the $O(N)$ symmetry stays unbroken the singlets $S_i$ are stable and are 
suitable as candidates for the dark matter in the 
universe~\cite{dm0,dm1,dm2,dm3,dm4}.

To be more concrete let us discuss the Lagrangian of the
model,  containing the standard model
Higgs boson plus an O(N)-symmetric sigma model.
The Lagrangian density is the following:

\begin{equation} 
L_{Scalar} = L_{Higgs} + L_{S} + L_{Interaction}     
\end{equation}
\begin{equation}
L_{Higgs}  = 
 -\frac{1}{2} D_{\mu}\Phi^{\dagger} D_{\mu}\Phi -{\lambda \over 8} \,
 (\Phi^{\dagger}\Phi - f^2)^2 
\end{equation}
\begin{equation}
L_{S}  = - \frac{1}{2}\,\partial_{\mu} \vec S \, 
\partial_{\mu}\vec S
     -\frac{1}{2} m_{S}^2 \,\vec S^2 - \frac{\lambda_S}{8N} \, 
     (\vec S^2 )^2 
\end{equation}
\begin{equation}
L_{Interaction} = -\frac{\omega}{4\sqrt{N}}\, \, \vec S^2 \,\Phi^{\dagger}\Phi 
\end{equation}  

The field $\Phi=(\sigma+f+i\pi_1,\pi_2+i\pi_3)$ is the complex Higgs doublet of 
the standard model with the
vacuum expectation value $<0|\Phi|0> = (f,0)$, $f=246$ GeV. Here, 
$\sigma$ is the physical  
Higgs boson and $\pi_{i=1,2,3}$ are the three Goldstone bosons. 
$\vec S = (S_1,\dots,S_N)$ is a real vector with \hbox{$<0|\vec S|0>= \vec 0$.} 
We consider the case, where the $O(N)$ symmetry stays unbroken,
because we want to concentrate on the effects of a finite width
of the Higgs particle. Breaking the $O(N)$ symmetry would lead
to more than one Higgs particle, through mixing. After the
spontaneous breaking of the standard model gauge symmetry the $\pi$ fields
become the longitudinal polarizations of the vector bosons.
In the unitary gauge one can simply put them to zero. One
is then left with an additional interaction in the Lagrangian
of the form: 

\begin{equation}
L_{Interaction} = -\frac{\omega f}{2\sqrt{N}}\, \, \vec S^2 \,\sigma 
\end{equation}  
This interaction leads to a decay into the $\vec S$ particles, that
do not couple to other fields of the standard model Lagrangian. On has therefore
an invisible width:
\begin{equation}
\Gamma_{Higgs}(invisible) = \frac {\omega^2}{32 \pi}\, \, 
\frac {f^2}{m_{Higgs}} 
(1-4 m_S^2/m_{Higgs}^2)^{1/2}
\end{equation}
This width is larger than the standard model width even for moderate values of $\omega$,
because the standard model width is strongly suppresed by the Yukawa coupings of the
fermions. Therefore the Higgs boson decays predominantly
invisibly with a branching ratio approximating 100\%.
Moreover one cannot exclude a large value of $\omega$.
In this case the Higgs is wide and decaying invisibly. 
This explains the name stealth model for this kind of Higgs sector.

However, is this Higgs boson undetectable at the LHC?
Its production mechanisms are exactly the same as the standard model 
ones, only its decay is in undetectable particles. One therefore
has to study associated production with an extra Z-boson or one must
consider the vector-boson fusion channel with jet-tagging. Assuming
the invisible branching ratio to be large and assuming the Higgs boson
not to be heavy, as indicated by the precision tests, one still finds
a significant signal~\cite{ekelof}. Of course one cannot study
this Higgs boson in great detail at the LHC. For this the ILC
would be needed, where precise measurements are possible in the
channel $e^+e^-\rightarrow Z H$.

\subsection{Mixing: fractional Higgses}

Somewhat surprisingly it is possible to have a model
that has basically only singlet-doublet mixing even if
all the scalars are light. If one starts with an interaction of the form
$H \Phi^{\dagger}\Phi$, where H is the new singlet Higgs field and $\Phi$ the standard model
Higgs field, no interaction of the form $H^3$, $H^4$ 
or $H^2 \Phi^{\dagger}\Phi$ is generated with an infinite
coefficient~\cite{hill}. 
At the same time the scalar potential stays bounded from below.
This means that one can indeed leave these dimension four interactions out of the Lagrangian
without violating renormalizability. This is similar to the non-renormalization theorem
in supersymmetry that says that the superpotential does not get renormalized.
However in general it only works with singlet extensions. As far as the counting of parameters
is concerned this is the most minimal extension of the standard model, having
only two extra parameters.

 The simplest model is the Hill model: 
\begin{equation}
L = -\frac{1}{2}(D_{\mu} \Phi)^{\dagger}(D_{\mu} \Phi) 
-\frac {1}{2}(\partial_{\mu} H)^2 - \frac {\lambda_0}{8}
(\Phi^{\dagger} \Phi -f_0^2)^2  -
\frac {\lambda_1}{8}(2f_1 H-\Phi^{\dagger}\Phi)^2 
\end{equation}
Working in the unitary gauge one writes $\Phi^{\dagger}=(\sigma,0)$,
where the $\sigma$-field is the physical standard model Higgs field.
Both the standard model Higgs field $\sigma$ and the Hill field $H$ receive vacuum expectation
values and one ends up with a two-by-two mass matrix to diagonalize, thereby
ending with two masses $m_-$ and $m_+$ and a mixing angle $\alpha$. There are two
equivalent ways to describe this situation. One is to say that one has two Higgs
fields with reduced couplings g to standard model particles:
\begin{equation}
g_-= g_{SM} \cos(\alpha), \qquad g_+= g_{SM} \sin(\alpha)
\end{equation} 
Because these two particles have the quantum numbers of the Higgs particle,
but only reduced couplings to standard model particles one can call them fractional Higgs
particles.
The other description, which has some practical advantages is not to diagonalize
the propagator, but simply keep the $\sigma - \sigma$ propagator explicitely.
One can ignore the $H-\sigma$ and $H-H$ propagators, since the $H$ field does not
couple to ordinary matter. One simply replaces in all experimental cross section
calculations the standard model Higgs propagator by:
\begin{equation}
D_{\sigma\sigma} (k^2) = \cos^2(\alpha)/(k^2 + m_-^2) + \sin^2(\alpha)/(k^2 + m_+^2)
\end{equation}
The generalization to an arbitrary set of fields $H_k$ is straightforward, one 
simply replaces the singlet-doublet interaction term by:
\begin{equation}
L_{H \Phi}= - \sum \frac {\lambda_k}{8}(2f_k H_k-\Phi^{\dagger}\Phi)^2 
\end{equation}
This will lead to a number of (fractional) Higgs bosons $H_i$ with reduced
couplings $g_i$ to the standard model particles such that
\be \sum_i g_i^2 = g^2_{SM} \ee

\subsection{A higher dimensional Higgs boson}
The mechanism described above can be generalized to an infinite number
of Higgses. The physical Higgs propagator is then given by an infinite number
of very small Higgs peaks, that cannot be resolved by the detector.
Ultimately one can take a continuum limit,
so as to produce an arbitray line shape for the Higgs boson, satisfying
the K\"all\'en-Lehmann representation.

\be D_{\sigma \sigma}(k^2)= \int ds\, \rho(s)/(k^2 + \rho(s) -i\epsilon) \ee

One has the sum rule~\cite{akhoury,gunion}
 $\int \rho(s)\, ds = 1$, while otherwise the theory is not renormalizable
and would lead to infinite effects for instance on the LEP precision variables.
Moreover, combining mixing with invisible
decay,  one can vary the invisible decay branching ratio
as a function of the invariant mass inside the Higgs propagator.
There is then no Higgs peak to be found any more.
The general Higgs propagator for the Higgs boson in the presence of
singlet fields is therefore determined by two function, the \KL\, spectral density
and the s-dependent invisible branching ratio. Unchanged compared to the
standard model are the relative branching ratio's to standard model particles.

Given the fact that the search for the
Higgs boson in the low mass range heavily depends on the presence of
a sharp mass peak, this is a promising way to hide the Higgs boson at the LHC.
However the general case is rather arbitrary and unelegant and ultimately
involves an infinite number of coupling constants. The question is therefore
whether there is a more esthetic way to generate such a spread-out Higgs 
signal, without the need of a large number of parameters. Actually this is 
possible. Because the $H \Phi^{\dagger}\Phi$ interaction
is superrenormalizable one can  let the $H$ field move in more
dimensions than four, without violating renormalizability. 
One can go up to six dimensions. The precise form of the propagator
will in general depend on the size and shape of the higher dimensions.
The exact formulas can be quite complicated. However it is possible that
these higher dimensions are simply open and flat. In this case one finds
simple formulas. One has for the generic case a propagator of the form:

\be
D_{\sigma \sigma}(q^2)= \left[ q^2 +M^2 - \mu_{lhd}^{8-d}
(q^2+m^2)^{d-6 \over 2} \right]^{-1} .\ee

For six dimensions one needs a limiting procedure and finds:

\be
D_{\sigma \sigma}(q^2)= \left[ q^2 +M^2 +\mu_{lhd}^2\,
\log(\frac{q^2+m^2}{\mu_{lhd}^2}) \right]^{-1} .
\ee

The parameter $M$ is a four-dimensional mass, $m$ a higher-dimensional mass
and $\mu_{lhd}$ a higher-to-lower dimensional mixing mass scale.
When one calculates the corresponding \KL\, spectral densities one finds
a low mass peak and a continuum that starts a bit higher in the mass.
The location of the peak is given by the zero of the inverse propagator.
Because this peak should not be a tachyon, there is a constraint on
$M, m, \mu_{lhd}$, that can be interpreted as the condition that there
is a stable vacuum. 

Explicitely one finds for $d=5$ the K\"all\'en-Lehmann spectral density:
\begin{eqnarray}
\rho(s) = &\theta(m^2-s)\,\, \frac{2(m^2-s_{peak})^{3/2}}{2(m^2-s_{peak})^{3/2}
+\mu_{lhd}^3}
\,\,\delta(s-s_{peak}) \nonumber\\   
+& \frac{\theta(s-m^2)}{\pi}\,\,
\frac{\mu_{lhd}^3\,(s-m^2)^{1/2}}{(s-m^2)(s-M^2)^2+\mu_{lhd}^6} ,
\end{eqnarray}

For $d=6$ one finds:
\begin{eqnarray}
\rho(s) = &\theta(m^2-s)\,\, \frac{m^2-s_{peak}}{m^2+\mu_{lhd}^2-s_{peak}}
\,\,\delta(s-s_{peak}) \nonumber\\
+& \theta(s-m^2)\,\,\frac{\mu_{lhd}^2}
{[\,s-M^2-\mu_{lhd}^2\,\log((s-m^2)/\mu_{lhd}^2)\,]^2+\pi^2\,\mu_{lhd}^4} .
\end{eqnarray}

If one does not introduce further fields no invisible decay is present.
If the delta peak is small enough it will be too insignificant for the LHC search.
The continuum is in any case difficult to see. There might possibly be a few sigma
signal in the $\tau$-sector. However if one adds to this model some scalars
to account for the dark matter, this will water down any remnant signal to
insignificance.

\section{Comparison with the LEP-200 data}
We now confront the higher dimensional models with the results from the direct
Higgs search at LEP-200~\cite{lep200}.
Within the pure standard model the absence of a clear signal has led to
a lower limit on the Higgs boson mass of $114.4 \gev$ at the 95\% confidence level.
Although no clear signal was found the data have some intriguing features,
that can be interpreted as evidence for Higgs bosons beyond the standard
model. There is a $2.3\,\sigma$ effect seen by all experiments at around 98 GeV.
A somewhat less significant $1.7\,\sigma$ excess is seen around 115 GeV. Finally
over the whole range $s^{1/2} > 100\gev$ the confidence level is less than
expected from background.
We will interpet these features as evidence for a spread-out Higgs-boson~\cite{dilcher}.
The peak at $98 \gev$ will be taken to correspond to the delta peak in the
\KL\, density. The other excess data will be taken as part of the continuum,
that will peak around $115 \gev$.

We start with the case $d=5$.
The delta-peak will be assumed to correspond to the
peak at 98 GeV, with a fixed value of $g^2_{98}$.
Ultimately we will vary the location of the peak between
$95\gev < m_{peak} < 101\gev$ and $0.056 < g^2_{98} < 0.144$.
After fixing $g^2_{98}$ and $m_{peak}$ we have one free variable,
 which we take to be $\mu_{lhd}$. If we also take a fixed
value for $\mu_{lhd}$ all parameters and thereby the
spectral density is known. We can then numerically integrate the
spectral density over selected ranges of $s$. The allowed range of $\mu_{lhd}$ 
is subsequently determined by the data at 115 GeV.
Since the peak at 115 GeV is not very well constrained, we
demand here only that the integrated spectral density
from $s_{down} = (110\gev)^2$ to $s_{up} = (120\gev)^2$
is larger than 30\%. This condition, together with formula (15),
which implies:
\be \rho(s) < \frac{(s-m^2)^{1/2}}{\pi\,\mu_{lhd}^3} , \ee

leads to the important analytical result:
\be 
\frac{2}{3\pi\,\mu_{lhd}^3} [\,(s_{up}-m_{peak}^2)^{3/2} - 
(s_{down}-m_{peak}^2)^{3/2}\,]
>0.3 \ee
This implies $\mu_{lhd} < 53\gev$. Using the constraint from
the strength of the delta-peak, it follows that the continuum starts
very close to the peak, the difference being less than 2.5 GeV.
This allows for a natural explanation, why the CL for the fit in the
whole range from 100 GeV to 110 GeV is somewhat less than what is expected by
pure background. The enhancement can be due to a slight, spread-out Higgs signal.
Actually when fitting the data with the above conditions, one finds for small 
values of $\mu_{lhd}$, that the integrated spectral density in the range
100 GeV to 110 GeV can become rather large, which would lead to problems
with the 95\% CL limits in this range. We therefore additionally demand
that the integrated spectral density in this range is less than 30\%.
There is no problem fitting the data with these conditions. As allowed ranges
we find:
\begin{eqnarray}
& 95\gev< m < 101\gev \nonumber\\
& 111\gev< M < 121\gev \nonumber\\
& 26\gev < \mu_{lhd} < 49\gev 
\end{eqnarray}

We now repeat the analysis for the case $d=6$.
The analytic argument gives the result:
\be
\frac{s_{up}-s_{down}}{\pi^2\,\mu_{lhd}^2} > 0.3
\ee
which implies $\mu_{lhd}<28\gev$.
Because of this low value of $\mu_{lhd}$ it is difficult
to get enough spectral weight arond 115 GeV and one also tends to get
too much density below 110 GeV. As a consequence the fit was only possible in a 
restricted range. Though not quite ruled out, the six-dimensional
case therefore seems to be somewhat disfavoured compared to the five-dimensional
case. As a consequence the fit was only possible in a restricted range.
We found the following limits:

\begin{eqnarray}
& 95\gev< m < 101\gev \nonumber\\
& 106\gev < M < 111\gev \nonumber\\
& 22\gev < \mu_{lhd} < 27\gev 
\end{eqnarray}

\section{Conclusion}
We are now in a position to answer the following question.
Is it possible to have a simple model that:\\

\noindent a) Is consistent with the precision data, even with the strong condition
  $m_H < 109 \gev$~?\\
b) explains the LEP-200 Higgs search data~?\\
c) has a dark matter candidate~?\\
d) gives no Higgs signal at the LHC~?\\

Given the above discussion, the answer is clearly yes, which 
leads to the question whether
such a model is likely to be true. This is rather difficult to answer
decisively. It depends on how significant the evidence in the data is,
in particular in the LEP-200 Higgs search data.
This significance is hard to estimate, since the data were not analyzed with
this type of model in mind. Taking the situation at face value the
spread-out singlet models
appear to be the only way to satisfy the experimental constraints.
In that case one is led to the conclusion that the LHC will not see a signal
for the Higgs boson.

\section*{Acknowledgments}
This work was supported by the BMBF Schwerpunktsprogramm
"Struktur und Wechselwirkung fundamentaler Teilchen".

\end{document}